\documentclass[twoside]{article}
\usepackage{latexsym,amsmath,amssymb}
\usepackage{hyperref}
\usepackage{cite} 
\usepackage{sectsty}
  \sectionfont{\large} \subsectionfont{\normalsize}
\usepackage{graphicx}

\usepackage{theorem}
\usepackage[all]{xy}
\newtheorem{theorem}{Theorem}[section]
{\theorembodyfont\rmfamily

}

\renewenvironment{abstract}
 {\small\begin{quote}{\textbf{Abstract}}\,\,}{\end{quote}}
\newenvironment{keywords}
 {\small\begin{quote}{\textbf{Keywords}}\,\,}{\end{quote}}
\newenvironment{classification}
 {\small\begin{quote}{\textbf{2010 Mathematics Subject Classification}}\,\,}{\end{quote}}
\date{}

\title{\vspace{-9ex}
{\centering
 \textbf{\large Numerical analysis for a unified 2 factor model of structural and reduced form types for corporate bonds with fixed discrete coupon
}}}

 \vspace{-4ex}

\author{\small\textsf{\bfseries 
$^{1}$ Hyong-chol O, $^{2}$Jong-Chol Kim, $^{2}$Il-Gwang Jon}\\[-.5ex]
{\footnotesize  ${}^{1, 2}$ Faculty of Mathematics, \textbf{Kim Il Sung} University,}
{\footnotesize   Pyongyang , D P R K}\\[-.5ex]
{\footnotesize e-mail: $^{1}$hc.o@ryongnamsan.edu.kp }}

\begin{document}

\maketitle
\thispagestyle{empty}

\vspace{-.6cm}

\begin{abstract}
Conditions of Stability for explicit finite difference scheme and some results of numerical analysis for a unified 2 factor model of structural and reduced form types for corporate bonds with fixed discrete coupon are provided. It seems to be difficult to get solution formula for PDE model which generalizes Agliardi’s structural model \cite{AA}  for discrete coupon bonds into a unified 2 factor model of structural and reduced form types and we study a numerical analysis for it by explicit finite difference scheme. These equations are parabolic equations with 3 variables and they include mixed derivatives, so the explicit finite difference scheme is not stable in general. We find conditions for the explicit finite difference scheme to be stable, in the case that it is stable, numerically compute the price of the bond and analyze its credit spread and duration.
\end{abstract}



%
%

\section{Introduction}

\indent

The study on defaultable corporate bonds and credit risk is now one of the most promising areas of cutting edge in financial mathematics \cite{AA}. As well known, there are two main approaches to pricing defaultable corporate bonds; one is the structural approach and the other one is the reduced form approach. In the structural method, we think that the default event occurs when the firm value is not enough to repay debt, that is, the firm value reaches a certain lower threshold (default barrier) from the above. Such a default can be expected and thus we call it expected default \cite{LS,M}. In the reduced-form approach, the default is treated as an unpredictable event governed by a default intensity process. In this case, the default event can occur without any correlation with the firm value and such a default is called unexpected default. In the reduced-form approach, if the default probability in time interval    is  , then   is called default intensity or hazard rate \cite{DS,J,JT,JLT}. The third approach is to unify the structural and reduced form approaches\cite{BP1,BP2,BSW,BYB,CL1,CL2,MU1,MU2,ODS,OM,OYD,ON,R}. As for the history of the above three approaches and their advantages and shortcomings, readers can refer to \cite{LS} and the introductions of \cite{BP1,CL1,CL2,ON}. Combining the elements of the structural approach and reduced-form approach is one of the recent trends \cite{CL1}.

On the other hand, the information including default barrier or default intensity is related to the firm’s financial data and it is difficult for investors outside of the firm to know the firm’s financial data in the whole lifetime of the bond. Investors outside of the firm could only know the management data discretely announced once in a certain term (for example, every year or every quarter). Therefore in 2005, Jiang’s group in Tongji university proposed the problem of pricing corporate bond using these discrete default information. Their aim is to close the study for credit risk to financial reality. In this direction, some results on defaultable zero coupon bonds are provided in \cite{ODS,OYD} using higher order binaries (\cite{OM}).

While there has been an enormous amount of theoretical modelling for zero-coupon bond pricing, starting from Merton \cite{M}, there has been relatively little work on the most realistic payout structure providing fixed discrete coupons \cite{AA}. Geske (1977) is the first study for this problem, where discrete interest payouts prior to maturity were modeled as determinants of default risk [9]. On the other hand, many models related to coupon approximate actual coupon bearing debts with continuous coupon stream or even zero coupon contracts but such approach has restriction \cite{AA,J}. The introduction and the conclusions of \cite{AA} include useful information about corporate discrete coupon bonds. Recently, Agliardi (2011) generalized the Geske’s formula for defaultable coupon bonds, incorporated a stochastic risk free term structure and the effects of bankruptcy cost and government taxes on bond interest and studied the duration of defaultable bonds. Agliardi’s approach in \cite{AA} to corporate coupon bonds is a kind of structural approaches as shown in its title. 

In \cite{ODC,OJS} authors studied the problem to generalize the results of \cite{AA} into a unified model of structural and reduced form models. To do this, in \cite{OJJ}, authors studied some general properties (including monotonicity and gradient estimates) of solutions to inhomogeneous Black-Scholes equations with discontinuous coefficients. The pricing formula of unified one factor model for fixed discrete coupon bond is provided in \cite{OJS}. It seems to be difficult to get the pricing formula of unified two factor model for fixed discrete coupon bond. Thus authors provide a unified two factor model for discounted discrete coupon bond and its pricing formula in \cite{ODC}. We believe that Agliardi’s structural 2 factor model for fixed discrete coupon bond has no analytical formula.

In this article we provide a unified two factor model of structural and reduced form types for corporate bonds with fixed discrete coupon which is a set of initial value problems of PDEs and numerical analysis for it by explicit finite difference scheme. These equations are parabolic equations with 3 variables and they include mixed derivatives, so the explicit finite difference scheme is not stable in general. We find conditions for the explicit finite difference scheme to be stable, in the case that it is stable, numerically compute the price of the bond and analyze its credit spread and duration.

\section{A unified 2 factor model for corporate bonds with fixed discrete coupon and its analysis}

\indent

{\bf Asumption}

1) Under the risk neutral martingale measure and a standard Wiener process $W_1$, the short rate follows the Vasicek model $dr_t=(a_1-a_2)dt+S_rdW_1(t)$. Under this assumption, the price $Z(r,t;T)$ of default free zero coupon bond with maturity T and face value 1 is the solution to the following problem

\begin{equation}\label{eq2-1}
\frac{\partial Z}{\partial t}+\frac{1}{2}S_r^2(t)\frac{\partial^2 Z}{\partial r^2}+a_r(r,t)-rZ=0,~
Z(r,T;T)=1.\tag{2.1}
\end{equation}

The solution is given by

\begin{equation}\label{eq2-2}
Z(r,t;T)=e^{\tilde{A}(r,T)-\tilde{B}(t,T)r}.\tag{2.2}
\end{equation}

\begin{equation}\label{eq2-3}
\tilde{B}(t,T)=\frac{1-e^{a_2(T-t)}}{a_2},~\tilde{A}(t,T)=-\int_t^T \left[a_1\tilde{B}(u,T)-\frac{1}{2}S_r^2\tilde{B}^2(u,T)\right]du.\tag{2.3}
\end{equation}

Here $\tilde{A}(t,T)$ and $\tilde{B}(t,T)$ are respectively given as follows:

\indent
2) The firm value  follows a geometric Brown motion 

$$dV(t)=(r_t-b)V(t)dt+S_V(t)V(t)dW_2(t)$$

under the risk neutral martingale measure and a standard Wiener process $W_2$ and $E(dW_1,dW_2)=\rho dt$. The firm continuously pays out dividend in rate $b\ge 0$  (constant) for a unit of firm value.

\indent
3) Let $0=T_0<T_1<\cdots<T_{N-1}<T_N=T$ and $T$ is the maturity of our corporate bond with face value  (unit of currency). At time $T_1(i=1,\cdots,N-1)$ bond holder receives the prior coupon of quantity $C_i$ (unit of currency) from the firm (this type of coupon is \textit{fixed} coupon) and at time $T_N=T$ bond holder receives the face value $F$ and the last coupon $C_N$ (unit of currency). 

\indent
4) The expected default occurs only at time $T_i$ when the equity of the firm is not enough to pay debt and coupon. If the expected default occurs, the bond holder receives $\delta V$ \textit{as default recovery}. Here $0 \le \delta \le 1$ is called \textit{a fractional recovery rate} of firm value at default.

\indent
5) The unexpected default can occur at any time. The unexpected default probability in the time interval $[t,t+\Delta t]\cap[T_i,T_{i+1}]$ is $\lambda _i \Delta t(i=0,\cdots ,N-1)$. Here the default intensity  is a constant. If the unexpected default occurs at time $t\in (T_i,T_{i+1})$, the bond holder receives $\min\left\{\delta V,~F\cdot Z(r,t;T)+\sum_{k=i+1}^N C_k\cdot Z(r,t;T_k)\right\}$ as default recovery and the equity holder gets nothing. 

\indent
 6) In the subinterval$(T_i,T_{i+1}]$, the price of our corporate bond is given by a sufficiently smooth function $B_i(V,r,T)$.

Our problem here is to fine the bond pricing functions $B_i(V,r,t)$.

By using the methods in \cite{AA} , \cite{ODC},  we can know that the bond pricing functions $B_i(V,r,t)$ are the solutions to the following problems.

\begin{eqnarray*}
&&\frac{\partial B_i}{\partial t}+\frac{1}{2}\left[S_V^2V^2\frac{\partial ^2B_i}{\partial V^2}+2\rho S_V S_r V\frac{\partial ^2 B_i}{\partial V\partial r}+S_r^2\frac{\partial ^2B_i}{\partial r^2}\right]+(r-b)V\frac{\partial B_i}{\partial V}+a_r\frac{\partial B_i}{\partial r}
\\
&&-(r+\lambda _i)B_i+\lambda _i \min\{\delta V,\Phi _i(r,t)\}=0,~ T_i<t<T_{i+1},~V>0,~r>0,~i=0,\cdots,N-1
\end{eqnarray*}
\begin{equation}\label{eq2-4}
B_{N-1}(V,r,T_N)=(F+C_N)\cdot 1\{V\ge F+C_N\}+\delta V\cdot 1\{V<F+C_N\},\tag{2.4}
\end{equation}
\begin{eqnarray*}
B_i(V,r,T_{i+1})=[B_{i+1}(V,r,T_{i+1})+C_{i+1}]\cdot 1\{V\ge B_{i+1}(V,r,T_{i+1})+C_{i+1}\}+
\\
+\delta V\cdot 1\{V<B_{i+1}(V,r,T_{i+1})+C_{i+1}\},\qquad V>0,~r>0,~i=0,\cdots,N-2.
\end{eqnarray*}
Here
$$
a_r=a_1-a_2 r,~\Phi_i(r,t)=\sum_{k=i+1}^N C_k\cdot Z(r,t;T_k)+F\cdot Z(r,t;T_N).
$$
The family \eqref{eq2-4} of the solving problems of the PDEs is {\bf the unified two factor model} for corporate bond with {\bf fixed} discrete coupon.

\indent

{\bf Remark 1} In \cite{AA}, the case of $\lambda_i=0,~i=0,\cdots,N-1$ (structural model) was studied. In \cite{OJS} authors studied the unified {\it one} factor model of structural and reduced models (which is \eqref{eq2-4} with constant short rate, that is, in this case $S_r=a_r=0$ in \eqref{eq2-1} and $Z(t,T)=e^{-r(T-t)}$) and provided an analytical pricing formula. In \cite{ODC} authors studied the unified {\it two} factor model of structural and reduced models for the discounted discrete coupon bond (which is the problem \eqref{eq2-4} with $C_i Z(r,T_i;T_N)=1,~i=1,\cdots,N$ instead of $C_i=1,~i=1,\cdots,N$ and the inhomogeneous term $\left(\sum_{k=i+1}^N C_k+F\right)\cdot Z(r,t;T_N)$ instead of $\Phi_i(r,t)$) and provided an analytical pricing formula.

\indent

The model \eqref{eq2-4} does not seem to have the analytical pricing formula except for the case of $i=N-1$.
In the case of $i=N-1$, we have the same solution representation as in the case of $i=N-1$ in the formula (2.20) in \cite{ODC}.
\begin{equation}\label{eq2-5}
B_{N-1}(V,r,t)=Z(r,t;T_N)\cdot u_{N-1}(V/Z(r,t;T_N),t),~T_{N-1}<t<T_N,\tag{2.5}
\end{equation}
\begin{eqnarray*}
&&u_{N-1}(x,t)=
\\
&&=e^{-\lambda_{N-1}(T_N-t)}[(F+C_N)\cdot N_1(d^-(\frac{x}{F+C_N},t,T))+\delta\cdot x e^{-b(T-t)}\cdot N_1(-d^+(\frac{x}{F+C_N},t,T))]
\\
&&+\lambda_{N-1}\int_t^{T_N} e^{-\lambda_{N-1}(\tau-t)}[(F+C_N)\cdot N_1(d^-(\delta\cdot x/(F+C_N),t,\tau))+
\end{eqnarray*}
\begin{equation}\label{eq2-6}
+\delta\cdot x e^{-b(T-t)}N_1(-d^+(\delta\cdot /(F+C_N),t,\tau))]d\tau,~T_{N-1}\le t<T_N=T,~x>0.\tag{2.6}
\end{equation}
Here,
$$
N_1(x)=(\sqrt{2\pi})^{-1}\int_{-\infty}^x \exp[-y^2/2]dy,
$$
\begin{equation}\label{eq2-7}
d^\pm(x,t,T)=\left(\sqrt{\int_t^T \sigma^2(u)du}\right)^{-1}\left[\ln(x)-b(T-t)\pm\int_t^T \frac{\sigma^2(d)}{2}du\right],\tag{2.7}
\end{equation}
$$
\sigma^2(t)=S_V^2+2\rho S_V S_r \tilde{B}(t,T)+S_r^2\tilde{B}^2(t,T).
$$

In the case of $0\le i<N-1$, we can solve numerically by using the explicit difference scheme.

\section{Stability conditions for the explicit difference scheme of the pricing model of the fixed discrete coupon bonds.}

\indent

If we use the variable transformation $V=x$ in \eqref{eq2-4} and denote the pricing function with respect to $x$ just as $B_i$, then the PDE model is written as following.
$$
\frac{\partial B_i}{\partial t}+\frac{1}{2}\left[S_x^2\frac{\partial^2 B_i}{\partial x^2}+2\rho S_x S_r \frac{\partial^2 B_i}{\partial x \partial r}+S_r^2\frac{\partial^2 B_i}{\partial r^2}\right]+(r-b-S_x^2/2)\frac{\partial B_i}{\partial x}+a_r\frac{\partial B_i}{\partial r}-
$$
\begin{equation}\label{eq3-1}
-(r+\lambda)B_i+\lambda_i\cdot \min\{\delta e^x,~\Phi_i\}=0,\qquad T_i<t<T_{i+1},~x\in {\mathbf R},~r>0,\qquad\tag{3.1}
\end{equation}
\begin{eqnarray*}
B_i(x,r,T_{i+1})=[B_{i+1}(x,r,T_{i+1})+C_{i+1}]\cdot 1\{e^x\ge B_{i+1}(x,r,T_{i+1})+C_{i+1}\}+
\\
+\delta e^x\cdot 1\{e^x<B_{i+1}(x,r,T_{i+1})+C_{i+1}\},\qquad x\in{\mathbf R},~r>0,~i=0,\cdots,N-2.
\end{eqnarray*}
Here $a_1,~a_2,~S_x=S_V,~S_r$ are constants, $a_r=a_1-a_2 r$.

We construct a lattice in the region ${\mathbf R}\times {\mathbf R}\times (0,T)$ as $x_l,r_m,t_n$ and denote by $B(x_l,r_m,t_n) = B_{l,m}^n$. As usual, we use the forward difference for the time derivative and the central difference for the space derivative.
$$
\frac{\partial B}{\partial t}=\frac{B_{l,m}^n-B_{l,m}^{n-1}}{\Delta t},~\frac{\partial B}{\partial x}=\frac{B_{l+1,m}^n-B_{l-1,m}^n}{2\Delta x},~\frac{\partial^2 B}{\partial x^2}=\frac{B_{l+1,m}^n-2B_{l,m}^n+B_{l-1,m}^n}{\Delta x^2},
$$
$$
\frac{\partial B}{\partial r}=\frac{B_{l,m+1}^n-B_{l,m-1}^n}{2\Delta r},\qquad\frac{\partial^2 B}{\partial r^2}=\frac{B_{l,m+1}^n-B_{l,m}^n+B_{l,m-1}^n}{\Delta r^2}.
$$
For mixed derivatives, according to the purpose, we use 
\begin{equation}\label{eq3-2}
\frac{\partial^2 B}{\partial x\partial r}=\frac{B_{l+1,m+1}^n-B_{l+1,m-1}^n-B_{l-1,m+1}^n+B_{l-1,m-1}^n}{4\Delta x\Delta r}~\text{(central difference)}\tag{3.2}
\end{equation}
or
\begin{equation}\label{eq3-3}
\frac{\partial^2 B}{\partial x\partial r}=\frac{B_{l+1,m+1}^n-B_{l+1,m}^n-B_{l,m+1}^n+B_{l,m-1}^n}{\Delta x\Delta r}~\text{(forward difference)}\tag{3.3}
\end{equation}
or
\begin{equation}\label{eq3-4}
\frac{\partial^2 B}{\partial x\partial r}=\frac{B_{l,m+1}^n-B_{l,m}^n-B_{l-1,m+1}^n+B_{l-1,m}^n}{\Delta x\Delta r}\tag{3.4}
\end{equation}
In \eqref{eq3-4} we combine the forward difference with the backward difference. 

Using the central difference for mixed derivative, we can write the explicit difference scheme of \eqref{eq3-1} as following by 
\begin{eqnarray*}
&&\frac{B_{l,m}^n-B_{l,m}^{n-1}}{\Delta t}+\frac{1}{2}\left[S_x^2\frac{B_{l+1,m}^n-2B_{l,m}^n+B_{l-1,m}^n}{\Delta x^2}+\right.
\\
&&\left.+2\rho S_x S_r\frac{B_{l+1,m+1}^n-B_{l-1,m+1}^n-B_{l+1,m-1}^n+B_{l-1,m-1}^n}{4\Delta x\Delta r}+S_r^2\frac{B_{l,m+1}^n-2B_{l,m}^n+B_{l,m-1}^n}{\Delta r^2}\right]
\\
&&+\left(r[m]-b-\frac{S_x^2}{2}\right)\frac{B_{l+1,m}^n-B_{l-1,m}^n}{2\Delta x}+a_r\frac{B_{l,m+1}^n-B_{l,m-1}^n}{2\Delta r}-(r[m]+\lambda_i)B_{l,m}^{n-1}+
\end{eqnarray*}
\begin{equation}\label{eq3-5}
\qquad+\lambda_i\min\left\{\delta e^{x[l]},\sum_{k=i+1}^N C_k Z(r[m],t[n];T_k)+FZ(r[m],t[n];T_N)\right\}=0,~T_i\le t[n]\le T_{i+1}.\tag{3.5}
\end{equation}
 If we denote $\mu_x=\frac{S_x^2\Delta t}{\Delta x^2},~\mu_r=\frac{S_r^2 \Delta t}{\Delta r^2}$ in \eqref{eq3-5} and arrange this, we have
\begin{eqnarray*}
&&B_{l,m}^{n-1}(1+\Delta t(r[m]+\lambda_i))=(1-\mu_x-\mu_r)B_{l,m}^n+\left(\frac{1}{2}\mu_r+\frac{\Delta t a_r}{2\Delta r}\right)B_{l,m+1}^n+\left(\frac{1}{2}\mu_r-\frac{\Delta t a_r}{2\Delta r}\right)B_{l,m-1}^n
\\
&&+\frac{\rho}{2}\sqrt{\mu_x}\sqrt{\mu_r}B_{l+1,m+1}^n-\frac{\rho}{2}\sqrt{\mu_x}\sqrt{\mu_r}B_{l+1,m+1}^n+\left(\frac{1}{2}+\mu_x+\frac{\Delta t}{2\Delta x}\left(r[m]-b-\frac{S_x^2}{2}\right)\right)B_{l+1,m}^n
\\
&&+\left(\frac{1}{2}\mu_x-\frac{\Delta t}{2\Delta x}\left(r[m]-b-\frac{S_x^2}{2}\right)\right)B_{l-1,m}^n-\frac{\rho}{4}\sqrt{\mu_x}\sqrt{\mu_r}B_{l-1,m+1}^n+\frac{\rho}{4}\sqrt{\mu_x}\sqrt{\mu_r}B_{l-1,m-1}^n
\end{eqnarray*}
\begin{equation}\label{eq3-6}
+\Delta t\lambda_i\min\left\{\delta e^{x[l]},\sum_{k=i+1}^N C_kZ(r[m],t[n];T_k)+FZ(r[m],t[n];T_N)\right\}.\qquad\quad\tag{3.6}
\end{equation}
Consider the stability of this scheme. If we still denote the error of each step as $B_{\bullet,\bullet}^n$, we have 
\begin{eqnarray*}
B_{l,m}^{n-1}(1+\Delta t(r[m]+\lambda_i))=(1-\mu_x-\mu_r)B_{l,m}^n+\mu_r\left(\frac{1}{2}+\frac{a_r}{2S_r^2}\Delta r\right)B_{l,m+1}^n+\mu_r\left(\frac{1}{2}-\frac{a_r}{2S_r^2}\Delta r\right)B_{l,m-1}^n
\\
+\frac{\rho}{2}\sqrt{\mu_x}\sqrt{\mu_r}B_{l+1,m+1}^n-\frac{\rho}{2}\sqrt{\mu_x}\sqrt{\mu_r}B_{l+1,m-1}^n+\mu_x\left(\frac{1}{2}+\frac{\Delta x}{2S_x^2}\left(r[m]-b-\frac{S_x^2}{2}\right)\right)B_{l+1,m}^n\qquad\qquad
\end{eqnarray*}
\begin{equation}\label{eq3-7}
+\mu_x\left(\frac{1}{2}-\frac{\Delta x}{2S_x^2}\left(r[m]-b-\frac{S_x^2}{2}\right)\right)_B{l-1,m}^n-\frac{\rho}{4}\sqrt{\mu_x}\sqrt{\mu_r}B_{l-1,m+1}^n+\frac{\rho}{4}\sqrt{\mu_x}\sqrt{\mu_r}B_{l-1,m-1}^n.\tag{3.7}
\end{equation}
Thus if the conditions 
\begin{equation}\label{eq3-8}
\mu_x+\mu_r<1,~\frac{\Delta x}{2S_x^2}\left|r[m]-b-\frac{S_x^2}{2}\right|<\frac{1}{2},~\frac{|a_r|}{2S_r^2}\Delta r<\frac{1}{2}\tag{3.8}
\end{equation}
are satisfied,  then $\left|B_{l,m}^{n-1}\right|\le\frac{1+|\rho|\sqrt{\mu_x}\sqrt{\mu_r}}{1+\Delta t(r[m]+\lambda_i)}\varepsilon$ when $\left|B_{\bullet,\bullet}^n\right|\le\varepsilon$. Thus we've proved the following theorem.

\begin{theorem}\label{theorem3-1}
If $\rho=0$ and the assumption \eqref{eq3-8} holds,  then the explicit difference scheme \eqref{eq3-6} of the problem \eqref{eq3-1} is stable.
\end{theorem}

Now we discuss the case of $1\ge\rho>0$. By using the forward difference rate of mixed derivatives, we can rewrite the explicit scheme of the problem \eqref{eq3-1} as following.
\begin{eqnarray*}
&&\frac{B_{l,m}^n-B_{l,m}^{n-1}}{\Delta t}+\frac{1}{2}\left[S_x^2\frac{B_{l+1,m}^n-2B_{l,m}^n+B_{l-1,m}^n}{\Delta x^2}+\right.
\\
&&\left.+2\rho S_x S_r\frac{B_{l+1,m+1}^n-B_{l+1,m}^n-B_{l,m+1}^n+B_{l,m}^n}{\Delta x\Delta r}+S_r^2\frac{B_{l,m+1}^n-2B_{l,m}^n+B_{l,m-1}^n}{\Delta r^2}\right]+
\\
&&+\left(r[m]-b-\frac{S_x^2}{2}\right)\frac{B_{l+1,m}^n-B_{l-1,m}^n}{2\Delta x}+a_r\frac{B_{l,m+1}^n-B_{l,m-1}^n}{2\Delta r}-(r[m]+\lambda)B_{l,m}^{n-1}+
\\
&&+\lambda\min\left\{\delta e^{x[l]},\sum_{k=i+1}^N C_k Z(r[m],t[n];T_k)+FZ(r[m],t[n];T_N)\right\}=0,~T_i\le t[n]\le T_{i+1}.
\end{eqnarray*}
Just as before, let $\mu_x=\frac{S_x^2\Delta t}{\Delta x^2},~\mu_r=\frac{S_r^2\Delta t}{\Delta r^2}$ and arrange the above equation then we can get the following.
\begin{eqnarray*}
&&B_{l,m}^{n-1}(1+\Delta t(r[m]+\lambda))=
\\
&&=B_{l,m}^n+\frac{1}{2}\left[\mu_x(B_{l+1,m}^n-2B_{l,m}^n+B_{l-1,m}^n)+2\rho\sqrt{\mu_x}\sqrt{\mu_r}(B_{l+1,m+1}^n-B_{l+1,m}^n-B_{l,m+1}^n+\right.
\\
&&\left.+B_{l,m}^n)+\mu_r(B_{l,m+1}^n-2B_{l,m}^n+B_{l,m-1}^n)\right]+\frac{\Delta t}{2\Delta x}\left(r[m]-b-\frac{S_x^2}{2}\right)(B_{l+1,m}^n-B_{l-1,m}^n)+
\end{eqnarray*}
\begin{equation}\label{eq3-9}
~~~+\frac{\Delta t a_r}{2\Delta r}(B_{l,m+1}^n-B_{l,m-1}^n)+\Delta t\lambda\min\left\{\delta e^{x[l]},\sum_{k=i+1}^N C_k Z(r[m],t[n];T_k)+FZ(r[m],t[n];T_N)\right\}.\qquad\qquad\tag{3.9}
\end{equation}
The study of stability just needs the discussion of homogeneous equations. If we still denote the error of each step just as $B_{\bullet,\bullet}^n$, then we have
\begin{eqnarray*}
B_{l,m}^{n-1}(1+\Delta t(r[m]+\lambda))=B_{l,m}^n(1-\mu_x-\mu_r+\rho\sqrt{\mu_x}\sqrt{\mu_r})+\qquad
\\
+B_{l+1,m}^n\mu_x\left(\frac{1}{2}-\rho\sqrt{\frac{\mu_r}{\mu_x}}+\frac{\Delta x}{2S_x^2}\left(r[m]-b-\frac{S_x^2}{2}\right)\right)
\\
+B_{l-1,m}^n\mu_x\left(\frac{1}{2}-\frac{\Delta x}{2S_x^2}\left(r[m]-b-\frac{S_x^2}{2}\right)\right)+B_{l+1,m+1}^n\rho\sqrt{\mu_x}\sqrt{\mu_r}
\end{eqnarray*}
\begin{equation}\label{eq3-10}
+B_{l,m+1}^n\mu_r\left(\frac{1}{2}-\rho\sqrt{\frac{\mu_x}{\mu_r}}+\frac{a_r}{2S_r^2}\Delta r\right)+B_{l,m-1}^n\mu_r\left(\frac{1}{2}-\frac{a_r}{2S_r^2}\Delta r\right).\tag{3.10}
\end{equation}
The condition for the coefficients of terms($B_{l,m}^n$) of the right side not to be negative is as follows.
\begin{equation}\label{eq3-11}
\mu_x+\mu_r-\rho\sqrt{\mu_x}\sqrt{\mu_r}=(\sqrt{\mu_x}-\sqrt{\mu_r})^2+(2-\rho)\sqrt{\mu_x}\sqrt{\mu_r}\le 1,\tag{3.11}
\end{equation}
\begin{equation}\label{eq3-12}
\frac{1}{2}>\rho\sqrt{\frac{\mu_r}{\mu_x}},\qquad\frac{\Delta x}{2S_x^2}\left|r[m]-b-\frac{S_x^2}{2}\right|<\frac{1}{2}-\rho\sqrt{\frac{\mu_r}{\mu_x}},\tag{3.12}
\end{equation}
\begin{equation}\label{eq3-13}
\frac{1}{2}>\rho\frac{\sqrt{\mu_x}}{\sqrt{\mu_r}},\qquad\frac{\Delta r}{2S_r^2}|a_r|<\frac{1}{2}-\rho\sqrt{\frac{\mu_x}{\mu_r}},\tag{3.13}
\end{equation}
\begin{equation}\label{eq3-14}
\frac{\Delta x}{2S_x^2}\left|r[m]-b-\frac{S_x^2}{2}\right|<\frac{1}{2},\qquad\frac{\Delta r}{2S_r^2}|a_r|<\frac{1}{2}.\tag{3.14}
\end{equation}
If we can choose $\Delta t,~\Delta x,~\Delta r$ to satisfy these conditions then the scheme is stable because the sum of the coefficients in the right side of \eqref{eq3-10} is $1$.

The first equation of \eqref{eq3-14} is the special case of \eqref{eq3-12} and the second is the special case of \eqref{eq3-13}. To satisfy \eqref{eq3-12} and \eqref{eq3-13} at the same time, $1/(2\rho)>\sqrt{\mu_r/\mu_x}>2$ must hold and so we need $0<\rho<1/2$. For example, if $\rho=1/3$ then we need $2/3<\sqrt{\mu_r/\mu_x}<3/2$. Therefore we prove the following theorem.

\begin{theorem}\label{theorem3-2}
If $0<\rho<1/2$ and the assumptions \eqref{eq3-11}\eqref{eq3-12}\eqref{eq3-13} holds true then the explicit difference scheme \eqref{eq3-9} of the problem \eqref{eq3-1} is stable.
\end{theorem}

Finally let us see the case of $-1\le\rho<0$. By using the combination of the forward difference rate with backward difference rate for the mixed derivative, we can write the explicit difference scheme as following.
\begin{eqnarray*}
&&\frac{B_{l,m}^n-B_{l,m}^{n-1}}{\Delta t}+\frac{1}{2}\left[S_x^2\frac{B_{l+1,m}^n-2B_{l,m}^n+B_{l-1,m}^n}{\Delta x^2}+2\rho S_x S_r\frac{B_{l,m+1}^n-B_{l,m}^n-B_{l-1,m+1}^n+B_{l-1,m}^n}{\Delta x\Delta r}+\right.
\\
&&\left.+S_r^2\frac{B_{l,m+1}^n-2B_{l,m}^n+B_{l,m-1}^n}{\Delta r^2}\right]+\left(r[m]-b-\frac{S_x^2}{2}\right)\frac{B_{l+1,m}^n-B_{l-1,m}^n}{2\Delta x}+a_r\frac{B_{l,m+1}^n-B_{l,m-1}^n}{2\Delta r}
\\
&&-(r[m]+\lambda)B_{l,m}^{n-1}+\lambda\min\left\{\delta e^{x[l]},\sum_{k=i+1}^N C_kZ(r[m],t[n];T_k)+FZ(r[m],t[n];T_N)\right\}=0,
\\
&&\qquad\qquad\qquad\qquad\qquad\qquad ~T_i\le t[n]\le T_{i+1}.
\end{eqnarray*}
Let $\mu_x=\frac{S_x^2\Delta t}{\Delta x^2},~\mu_r=\frac{S_r^2\Delta t}{\Delta r^2}$ and arrange the above equation, we can get the following equation.
\begin{eqnarray*}
&&B_{l,m}^{n-1}[1+\Delta t(r[m]+\lambda)]=
\\
&&=B_{l,m}^n+\frac{1}{2}\left[\mu_x(B_{l+1,m}^n-2B_{l,m}^n+B_{l-1,m}^n)+2\rho\sqrt{\mu_x}\sqrt{\mu_r}(B_{l,m+1}^n-B_{l,m}^n-B_{l-1,m+1}^n+B_{l-1,m}^n)\right.
\\
&&\left.+\mu_r(B_{l,m+1}^n-2B_{l,m}^n+B_{l,m-1}^n)\right]+\frac{\Delta t}{2\Delta x}\left(r[m]-b-\frac{S_x^2}{2}\right)(B_{l+1,m}^n-B_{l-1,m}^n)+
\\
&&+\frac{\Delta t a_r}{2\Delta r}(B_{l,m+1}^n-B_{l,m-1}^n)+
\end{eqnarray*}
\begin{equation}\label{eq3-15}
+\Delta t\lambda\min\left\{\delta e^{x[l]},\sum_{k=i+1}^N C_kZ(r[m],t[n];T_k)+FZ(r[m],t[n];T_N)\right\}.\qquad\qquad\tag{3.15}
\end{equation}
The study of the stability can be done by discussing homogeneous equations. If we still denote the error of each step as $B_{\bullet,\bullet}^n$, then we have
\begin{eqnarray*}
&&B_{l,m}^{n-1}[1+\Delta t(r[m]+\lambda)]=B_{l,m}^n(1-\mu_x-\mu_r-\rho\sqrt{\mu_x}\sqrt{\mu_r})+
\\
&&+B_{l+1,m}^n\mu_x\left(\frac{1}{2}+\frac{\Delta x}{2S_x^2}\left(r[m]-b-\frac{S_x^2}{2}\right)\right)+
\\
&&+B_{l-1,m}^n\mu_x\left(\frac{1}{2}+\rho\sqrt{\frac{\mu_r}{\mu_x}}-\frac{\Delta x}{2S_x^2}\left(r[m]-b-\frac{S_x^2}{2}\right)\right)+
\\&&+B_{l,m+1}^n\mu_r\left(\frac{1}{2}+\rho\sqrt{\frac{\mu_x}{\mu_r}}+\frac{\Delta r}{2S_r^2}a_r\right)+B_{l,m+1}^n\mu_r\left(\frac{1}{2}-\frac{\Delta r}{2S_r^2}a_r\right)+
\end{eqnarray*}
\begin{equation}\label{eq3-16}
+B_{l-1,m+1}^n(-\rho\sqrt{\mu_x}\sqrt{\mu_r}).\qquad\qquad\qquad\qquad\qquad\qquad\qquad~~~\tag{3.16}
\end{equation}
The condition for the coefficients of every term $B_{l,m}^n$ of the right side of \eqref{eq3-16} not to be negative is 
\begin{equation}\label{eq3-17}
\mu_x+\mu_r+\rho\sqrt{\mu_x}\sqrt{\mu_r}=(\sqrt{\mu_x}-\sqrt{\mu_r})^2+(2+\rho)\sqrt{\mu_x}\sqrt{\mu_r}\le 1,\tag{3.17}
\end{equation}
\begin{equation}\label{eq3-18}
\frac{1}{2}>-\rho\sqrt{\frac{\mu_r}{\mu_x}},\qquad\frac{\Delta x}{2S_x^2}\left|r[m]-b-\frac{S_x^2}{2}\right|<\frac{1}{2}+\rho\sqrt{\frac{\mu_r}{\mu_x}},\tag{3.18}
\end{equation}
\begin{equation}\label{eq3-19}
\frac{1}{2}>-\rho\sqrt{\frac{\mu_x}{\mu_r}},\qquad\frac{\Delta r}{2S_r^2}|a_r|<\frac{1}{2}+\rho\sqrt{\frac{\mu_x}{\mu_r}},\tag{3.19}
\end{equation}
\begin{equation}\label{eq3-20}
\frac{\Delta x}{2S_x^2}\left|r[m]-b-\frac{S_x^2}{2}\right|<\frac{1}{2},\qquad\frac{\Delta r}{2S_r^2}|a_r|<\frac{1}{2}.\tag{3.20}
\end{equation}
If we can choose $\Delta t,~\Delta x,~\Delta r$ to satisfy these conditions the scheme is stable because the sum of the coefficients in the right side is $1$.
The first equation in \eqref{eq3-20} is the special case of \eqref{eq3-18} and the second equation is the special case of \eqref{eq3-19}. To satisfy \eqref{eq3-8} and \eqref{eq3-19} at the same time, $-1/(2\rho)>\sqrt{\mu_r/\mu_x}>-2\rho$ must hold true and so we need $2/3<\sqrt{\mu_r/\mu_x}<3/2$. Therefore we proved the following theorem. 
\begin{theorem}\label{theorem3-3}
If $0>\rho>-1/2$ and the assumption \eqref{eq3-17}\eqref{eq3-18}\eqref{eq3-19} holds true then the explicit differential scheme \eqref{eq3-15} of the problem \eqref{eq3-1} is stable.
\end{theorem}

\section{The result and analysis of the numerical study for fixed discrete coupon bond pricing model by the explicit difference scheme}

\indent

In this section we give the result of the numerical analysis about the fixed discrete coupon bond pricing model \eqref{eq2-4}. For simplification we only consider the case of $\rho=0$.

The basic data for computing are as following.
$$
N=2,~T_1=0.5,~T_2=1,~a_1=0.379\times 0.098,~a_2=0.379,~S_r=0.077,~b=0.05,
$$
$$
S_V=1.0,~\rho=0,~\lambda_0=0.1,~\lambda_1=0.3,~\delta=0.5,~F=10,~C_1=C_2=1.0.
$$
The lattice size of the explicit difference scheme is as following.
$$
\mu_x=0.0104,~\mu_r=0.9625,~\Delta t=0.005,~\Delta  r=0.02,~\Delta x=\ln 2.
$$

\subsection{The price of the bond}

\indent

Some of the computing results are as following.

\begin{center}
\begin{tabular}{|c|c|c|c|c|}
\hline
V/R & 0.02 & 0.04 & 0.06 & 0.08 \\ \cline{1-5}
5.00 & 3.257 & 3.251 & 3.245 & 3.238 \\ \cline{1-5}
9.11 & 5.103 & 5.081 & 5.059 & 5.037 \\ \cline{1-5}
10.06 & 5.452 & 5.427 & 5.401 & 5.375 \\ \cline{1-5}
11.13 & 5.812 & 5.782 & 5.751 & 5.721 \\ \cline{1-5}
12.30 & 6.178 & 6.143 & 6.108 & 6.073 \\ \cline{1-5}
13.60 & 6.549 & 6.509 & 6.469 & 6.428 \\ \cline{1-5}
20.30 & 8.028 & 7.961 & 7.894 & 7.826 \\ \cline{1-5}
30.20 & 9.333 & 9.233 & 9.134 & 9.034 \\ \cline{1-5}
33.40 & 9.612 & 9.504 & 9.397 & 9.290 \\ \cline{1-5}
\hline
\end{tabular}

Table 4.1  The initial ($t=0$) price of the bond corresponding to $V$ and $r$ 

\end{center}

\begin{center}
\begin{tabular}{|c|c|c|c|c|c|}
\hline
V/t & 0.00 & 0.25 & 0.50 & 0.75 & 1.00 \\ \cline{1-6}
5.00 & 3.251 & 3.485 & 3.606 & 2.599 & 2.500 \\ \cline{1-6}
9.11 & 5.081 & 5.465 & 5.845 & 5.297 & 4.555 \\ \cline{1-6}
10.06 & 5.427 & 5.852 & 6.302 & 5.900 & 5.030 \\ \cline{1-6}
11.13 & 5.782 & 6.251 & 6.782 & 6.533 & 11.000 \\ \cline{1-6}
12.30 & 6.143 & 6.658 & 7.264 & 7.162 & 11.000 \\ \cline{1-6}
13.60 & 6.509 & 7.068 & 7.750 & 7.776 & 11.000 \\ \cline{1-6}
20.30 & 7.961 & 8.661 & 9.540 & 9.736 & 11.000 \\ \cline{1-6}
30.20 & 9.233 & 9.959 & 10.769 & 10.623 & 11.000 \\ \cline{1-6}
33.40 & 9.504 & 10.215 & 10.979 & 10.719 & 11.000 \\ \cline{1-6}
\hline
\end{tabular}

 Table 4.2. The price of the bond corresponding to $V$ and $t$ when $r=0.04$
\end{center}

In what follows we give some graphical analysis based on the computing result of the bond price.

We can get the following conclusions by analyzing the graphs of the computing result.

Figure 4.1 provides the graphs of the initial price of bonds as $r$ changes from 0.02 to 0.1 in the case that $V$ is 5, 12.3, 33.4, respectively. From figure 4.1 we can know that the initial bond price decreases when the short rate increases. 

Figure 4.2 provides the graphs of  the  initial price of bonds as $V$ changes from 5 to 35 in the case that $r$ is 0.02, 0.04, 0.08, respectively. From figure 4.2 we can know that the initial bond price increases. when $V$ increases. And we can intuitively know that the bond pricing function is a convex function of the firm's value. 

Figure 4.3 provides the graphs of the bond price as $t$ changes in the case that $V=20.276$ and $r$ is 0.02, 0.04, 0.08, respectively. From figure 4.3 we can know that the bond price decreases in whole lifetime when the interest rate increases.

\begin{figure}[htbp]
\centerline{\includegraphics[width=3in,height=2.3in]{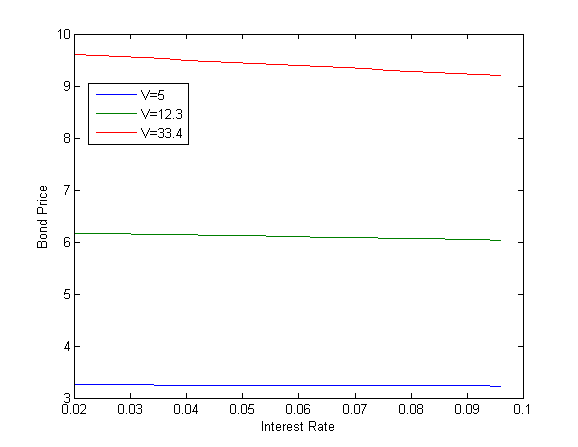}
\includegraphics[width=3in,height=2.3in]{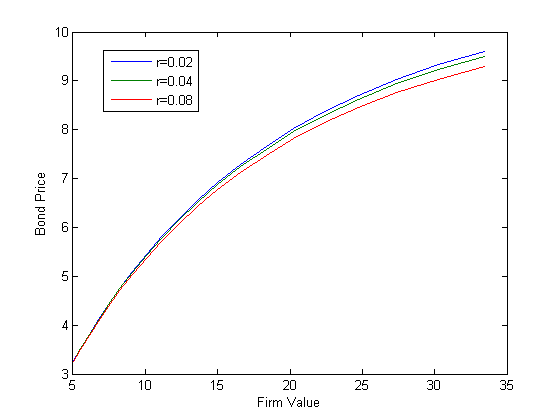}}
\centerline{{\small Figure-4.1 $r-B$ graph when $t=0$ \qquad\qquad\qquad\qquad\quad  Figure-4.2 $V-B$ graph when $t=0$}}
\vspace{1cm}

\centerline{\includegraphics[width=3in,height=2.3in]{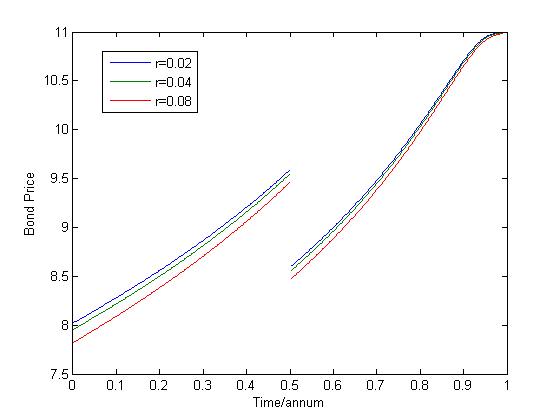}
\includegraphics[width=3in,height=2.3in]{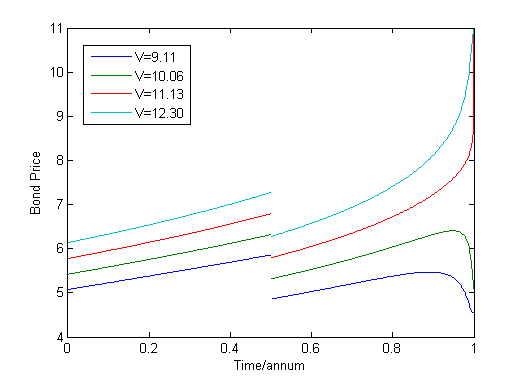}}
\centerline{{\small Figure-4.3 $t-B$ graph when $V=20.276$ \qquad\qquad\qquad\qquad\quad  Figure-4.4 $t-B$ graph when $r=0.04$}}
\end{figure}

Figure 4.4 provides the graphs of the bond price as $t$ changes in the case that $r=0.04$ and $V$ is 9.11, 10.06, 11.13, 12.3, respectively. From figure 4.4 we can know that the bond price rises when $V$ increases. And we can see that if the firm’s value is small, then the bond price falls since the possibility of the default event to occur becomes larger.

In figure 4.3 and 4.4 we can find that the bond price decreases with jumps on coupon dates. 

These results show that the numerical results on our model are compatible with the real financial situation.

\subsection{Credit spread}

\indent

If $C$ is a defaultable bond with maturity $T$, $Z$ is a risk free bond with the same maturity $T$, then we define the credit spread as follows:
$$
CS=-\frac{\ln{C(t)}-\ln{Z(t)}}{T-t}.
$$

If the short rate becomes larger, then the present value of the government bond becomes smaller than the future price. If there is a credit risk, then the defaultable bond price becomes smaller than the government bond. We can see as if this difference of prices is due to increasing the short rate just as the amount of existence of credit risk. It is just the idea of introduction of the credit spread. 

From the definition, we have $C(t)=Z(t)e^{-CS(T-t)}$. In the case of constant short rate, the price of risk free zero coupon bond is $Z(t)=e^{-r(T-t)}$ and thus we have $C(t)=e^{-(r+CS)(T-t)}$. So in this case we can see as if the short rate increases just as $CS$ if there is a credit risk. 

Now we compute the credit spread in the corporate bond pricing model \eqref{eq2-4} with fixed discrete coupons.

The present price of the government bond, the holder of which receives the prior coupon $C_1$  at the time $T_1$ and the last coupon $C_2$ and the face value $F$ at the time $T_2$, can be computed as follows:
\begin{equation}\label{eq4-1}
P(r,t:T_1,C_1,T_2,C_2)=\left\{
\begin{array}{rl}
(F+C_2)Z(r,t;T_2);\qquad\qquad\qquad T_1\le t\le T_2,
\\
(F+C_2)Z(r,t;T_2)+C_1Z(r,t;T_1);\quad 0\le t\le T_1.
\end{array}\right.\tag{4.1}
\end{equation}
Therefore, for the corporate bond, the holder of which receives the prior coupon $C_1$ at the time $T_1$ and the last coupon $C_2$ and the face value $F$ at the time $T_2$, the credit spread (added due to the credit risk) is computed as follows:
\begin{equation}\label{eq4-2}
P(r,t:T_1,C_1,T_2,C_2)=\left\{
\begin{array}{rl}
-\frac{\ln{\frac{B_1(V,r,t)}{(F+C_2)Z(r,t;T_2)}}}{T-t},\qquad\qquad\qquad T_1\le t\le T_2,
\\
-\frac{\ln{\frac{B_0(V,r,t)}{(F+C_2)Z(r,t;T_2)+C_1Z(r,t;T_1)}}}{T-t},\qquad~ 0\le t\le T_1.
\end{array}\right.\tag{4.2}
\end{equation}

In what follows, we give some graphical analysis based on the computing result.

\begin{figure}[htbp]
\centerline{\includegraphics[width=3in,height=2.3in]{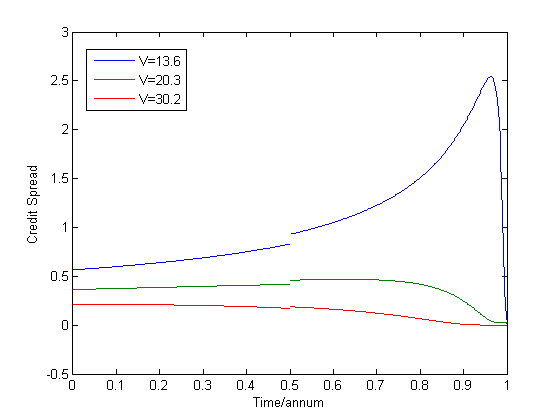}
\includegraphics[width=3in,height=2.3in]{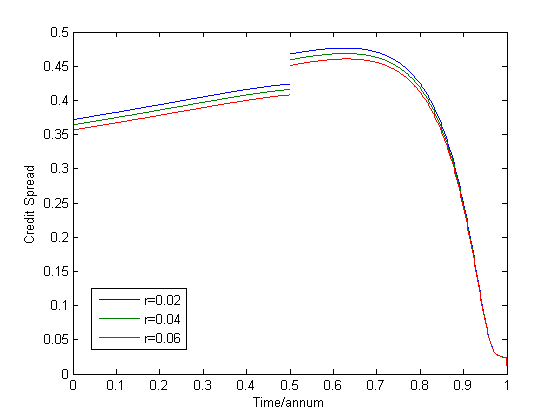}}
\centerline{{\small Figure-4.5 $t-CS$ graph when $r=0.04$ \qquad\qquad\qquad\quad  Figure-4.6 $t-CS$ graph when $V=20.3$}}
\end{figure}

From the resulting graphs we can get the following conclusions.

Figure 4.5 provides the graphs of the credit spread as  changes in the case that $V=13.6,~20.3,~30.2$, respectively. From the figure we can know that the credit spread decreases as the firm value increases and it increases rapidly as the time goes to the maturity date when the firm value is small.

Figure 4.6 gives the graphs of the credit spread as $t$ changes in the case that $r=0.02,~0.04,~0.06$, respectively. From the figure we can know that the credit spread decreases when $r$ increases. 

Figure 4.7 provides the graphs of the credit spread as  changes in the case that $V=20.276,~r=0.04$ and $C_1=C_2=0,~1,~2$ respectively. From the figure we can find that the credit spread curve doesn't have a jump in the case of zero coupon credit bond but the credit spread curve for coupon bond has a jump on the prior coupon date $T_1$, and the credit spread increases as the quantity of coupon increases.

Figure 4.8 gives the graphs of the time - credit spread in the case of different default intensities when $V=20.276,~r=0.04$. From the figure we can find that the credit spread increases when the default intensity increases. According to figure 4.5-4.8 the credit spread increases with a jump on the coupon date.

These results show that the numerical results on our model are compatible with the real financial situation.

\begin{figure}[htbp]
\centerline{\includegraphics[width=3in,height=2.3in]{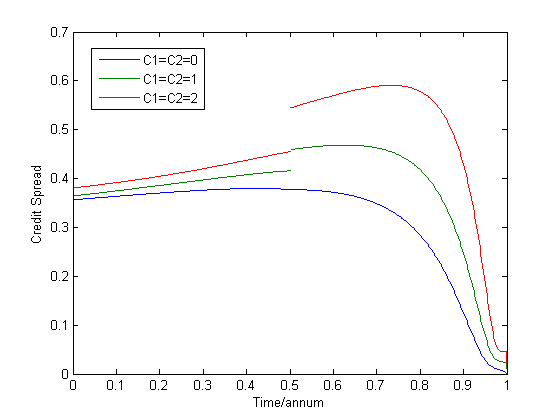}
\includegraphics[width=3in,height=2.3in]{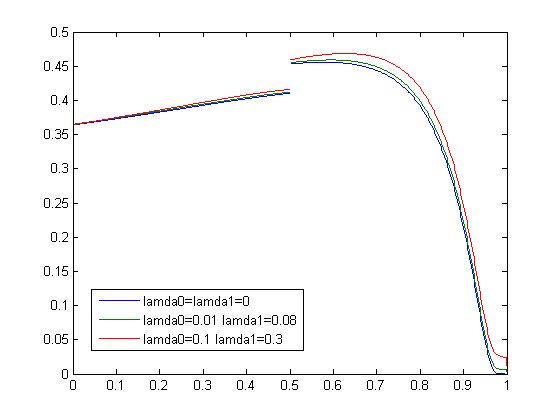}}
\centerline{{\small Figure-4.7 $t-CS$ graph when $V=20.3,~r=0.04$ \qquad  Figure-4.8 $t-CS$ graph when $V=20.3,~r=0.04$}}
\end{figure}

\subsection{Duration}

\indent

Let $B(V,r,t)$ be a corporate bond price. Then the {\bf duration} with respect to the short rate is defined as follows.
\begin{equation}\label{eq4-3}
D(V,r,t)=-\frac{1}{B(V,r,t)}\partial_r B(V,r,t).\tag{4.3}
\end{equation}
The duration of bond is a measure of how long on average the holder of the bond has to wait before receiving cash payments and it is an important concept in risk management \cite{H}.

When the short rate $r$ is constant, the price of zero coupon government bond with maturity $T$ (year) is $Z(t)=e^{-r(T-t)}$ and so $T-t$ is the duration. But in the case of coupon bearing bond or in the case that the short rate depends on time or some other factors, the situation becomes different.

When the short rate $r$ is constant and $0<t_1<\cdots<t_n$, the time $t(0<t<t_1)$ - price of the discrete coupon government bond that receives cash $C_i$ at every time  $t_i$($i=1,\cdots,n$) is $B(t)=\sum_{i=1}^n C_i e^{-r(t_i-t)}$ and from the definition, the duration is computed as follows:
\begin{equation}\label{eq4-4}
D=\frac{\sum_{i=1}^n (t_i-t)C_i e^{-r(t_i-t)}}{B}=\sum_{i=1}^n \frac{C_i e^{-r(t_i-t)}}{B(t)}(t_i-t).\tag{4.4}
\end{equation}
From $\sum_{i=1}^n \frac{C_i e^{-r(t_i-t)}}{B(t)}=1$, the duration is the {\it weighted average} of the {\it terms to the payment dates}. The weight indicates the ratio of the current price of the payment at time $t_i$ to the current price of the bond price, i. e, the {\it proportion} of the payment at time $t_i$ in the current price of the bond.

Figure 4.9 a gives the time - duration graph of coupon government bond when $n=2, C_1=1, C_2=1, r = 0.04$. When the interest rate is a random process following to the Vasicek model, the duration of zero coupon government bond whose price is given by \eqref{eq2-2} is $B(t,T)$. (See Figure 4-9 b for its graph.)
From figure 4.9 a) we can find that the duration increases after receiving coupon. 

\begin{figure}[htbp]
\centerline{\includegraphics[width=3in,height=2.3in]{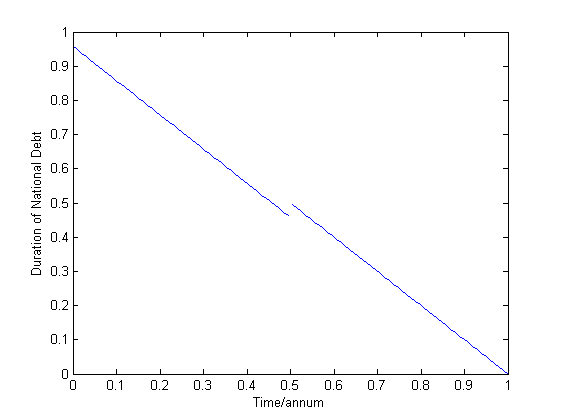}
\includegraphics[width=3in,height=2.3in]{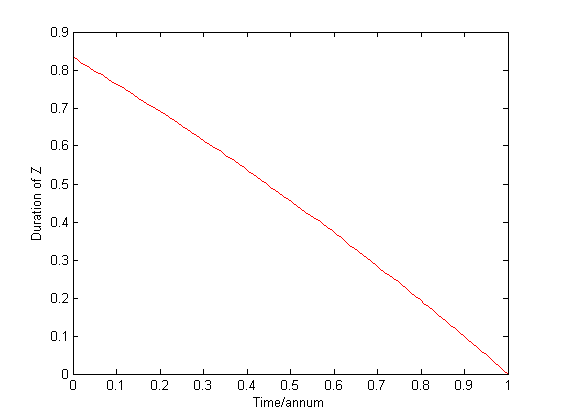}}
\centerline{{\small Figure-4.9 a) time - duration graph in \eqref{eq4-4} \qquad\qquad b) duration of defaultable coupon GB}}
\end{figure}

In what follows, we give some graphical analysis based on the computing result of the duration for fixed discrete coupon bond model \eqref{eq2-4}.

We can get the following conclusions by analyzing the resulting graph.

Figure 4.10 gives the time - graphs of the duration in the case that $r=0.04$ and $V$ is 13.6, 20.3, 30.2, respectively. From the figure we can find that the duration increases with respect to $V$. Figure 4.11 gives the time-graph of the duration in the case that $V=20.276$ and $r$ is 0.02, 0.04, 0.06, respectively. From the figure we can find that the duration increases with respect to $r$.

\begin{figure}[htbp]
\centerline{\includegraphics[width=3in,height=2.3in]{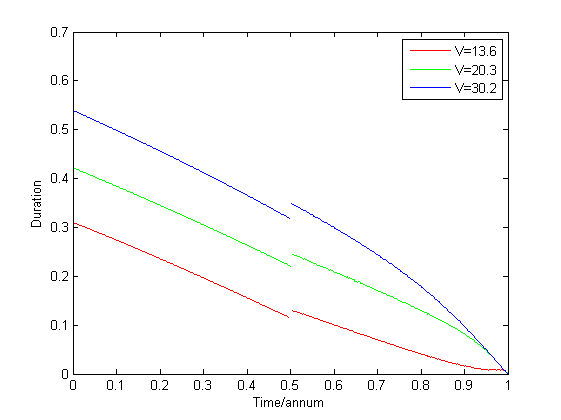}
\includegraphics[width=3in,height=2.3in]{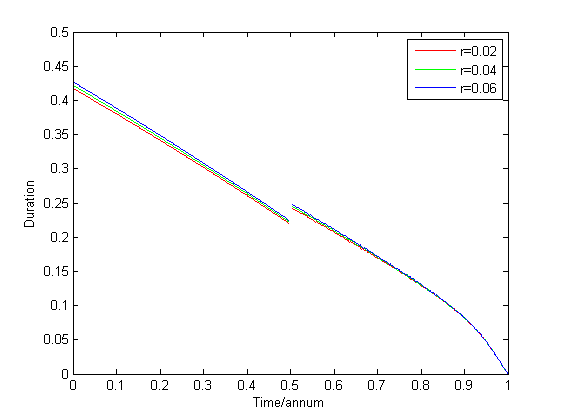}}
\centerline{{\small Figure-4.10 $t-D$ graph when $r=0.04$ \qquad\qquad Figure-4.11 $t-D$ graph when $V=20.3$}}
\end{figure}


\end{document}